\documentclass{spie}

\usepackage{cite}

\usepackage{times} % Times New Roman
\usepackage{indentfirst} % Indent first paragraphs

\usepackage[colorinlistoftodos,color=pink]{todonotes} % For fancy \todo{something}
\usepackage{xkeyval}
\presetkeys{todonotes}{inline}{}
\usepackage{dsfont}
\usepackage{float}
\usepackage{graphicx}
\usepackage{algorithmic}
\usepackage{algorithm}
\usepackage{amssymb}
\usepackage[utf8]{inputenc}
\usepackage{mathtools}
\usepackage{amsmath}
\usepackage{marvosym}
\usepackage{subcaption}
\usepackage{lipsum} % For lorem ipsum, remove this
\usepackage{tabularx}
\usepackage{comment}
\usepackage{caption}
\usepackage{hyperref}

\newcommand{\email}[1]{\texttt{Email: #1}}
\newcolumntype{C}{>{\centering\arraybackslash}X}

\title{Embracing Uncertainty Flexibility: Harnessing a Supervised Tree Kernel to Empower Ensemble Modelling for 2D Echocardiography-Based Prediction of Right Ventricular Volume}

\author{Tuan A. Bohoran \supit{1,*,$^{\textrm{\Letter}}$}, Polydoros N. Kampaktsis\supit{2,*}, Laura McLaughlin\supit{2}, 		
        \skiplinehalf
        Jay Leb\supit{2}, Serafeim Moustakidis\supit{3},
        Gerry P. McCann\supit{4},
        Archontis Giannakidis\supit{1}
  \skiplinehalf
  \normalsize 
  \supit{1}School of Science and Technology, Nottingham Trent University, Nottingham, UK.; \\
  \email{tuan.bohoran@ntu.ac.uk}; \\
  \supit{2}Division of Cardiology, Columbia University Irving Medical Center, New York City, NY, USA.; \\
  \supit{3}AiDEAS, Tallinn, Estonia.; \\
  \supit{4}Department of Cardiovascular Sciences, University of Leicester and the NIHR Leicester Biomedical Research Centre, Glenfield Hospital, Leicester, UK.;\\
  \supit{*} Authors contributed equally
}

\begin{document}

\maketitle

\begin{abstract}
The right ventricular (RV) function deterioration strongly predicts clinical outcomes in numerous circumstances.
To boost the clinical deployment of ensemble regression methods that quantify RV volumes using tabular data from the widely available two-dimensional echocardiography (2DE), we propose to complement the volume predictions with uncertainty scores.
To this end, we employ an instance-based method which uses the learned tree structure to identify the nearest training samples to a target instance and then uses a number of distribution types to more flexibly model the output.
The probabilistic and point-prediction performances of the proposed framework are evaluated on a relatively small-scale dataset, comprising 100 end-diastolic and end-systolic RV volumes. The reference values for point performance were obtained from MRI. 
The results demonstrate that our flexible approach yields improved probabilistic and point performances over other state-of-the art methods.
The appropriateness of the proposed framework is showcased by providing exemplar cases.
The estimated uncertainty embodies both aleatoric and epistemic types. This work aligns with trustworthy artificial intelligence since it can be used to enhance the decision-making process
and reduce risks. 
The feature importance scores of our framework can be exploited to reduce the number of required 2DE views which could enhance the proposed pipeline’s clinical application.

\end{abstract}

\section{Introduction}

Right ventricular systolic (RV) dysfunction is a powerful and independent mortality predictor \cite{Haddad2008} which may 
occur from a large variety of cardiovascular disorders that result in inability to pump enough blood for oxygenation. Machine learning methods have recently shown \cite{bohoran2023right} great potential in quantifying RV volumes using tabular data (such as area measurements, age, gender and cardiac phase information) obtained from the widely available and highly portable two-dimensional echocardiography (2DE). However, for clinical deployment where patient safety is at stake, it is crucial to complement these RV volume predictions with uncertainty scores that reflect the degree of trust in these predictions.

The goal of this paper is to present an uncertainty quantification framework when predicting RV volumes through the use of ensemble models, in particular Gradient-Boosted Regression Trees (GBRTs), on 2DE-derived tabular data. GBRTs are regarded \cite{NEURIPS2022_0378c769} the method of choice for tabular data. To get an estimate of the prediction uncertainty, we propose to make use of a $k$-nearest neighbour method \cite{Brophy2022Instance} that relies on a supervised tree kernel \cite{YiLin_2023,daghistani2020comparison}. Unlike other state-of-the-art (SOTA) gradient-boosted algorithms \cite{pmlr-v119-duan20a,Sprangers2021ProbabilisticGB,malinin2021uncertainty} that provide probabilistic predictions, this method performs well on both probabilistic and point performances, and can also use a number of distribution types to more flexibly model the output. It can also be applied to any GBRT model, adding further flexibility.

The probabilistic and point-prediction performances of our framework are evaluated on a relatively small-scale dataset, comprising 100 end-diastolic (ED) and end-systolic (ES) RV volumes. The reference values for point performance were obtained from cardiovascular MRI (CMR). Our pipeline is also compared to other SOTA methods.
Lastly, we provide conditional output distributions and the respective confidence intervals for a couple of high and low accuracy (test set) predictions.

\section{Materials \& Methods}

\subsection{Dataset}

The study population was a retrospective cohort of 50 adult patients for which 2DE and CMR were acquired. Data acquisition and annotation were as described in \cite{bohoran2023right}. In brief, for each patient the RV endocardial-myocardial interface was manually traced in end-systole and end-diastole (making a total of 100 data points) for the following eight standardised echocardiographic views: parasternal long axis (PLAX), right ventricular inflow (RV Inflow), parasternal short axis at the level of the aortic valve (PSAX AV), basal (PSAX Base), mid (PSAX mid) and apical left ventricular segments (PSAX Distal), four-chamber (Four C) and subcostal (Sub C) views. The eight area measurements along with the patient age were the numerical input variables of our model, whereas the gender and cardiac phase information were the categorical ones. The short-axis cine CMR-derived ED and ES RV volumes were recorded in a semi-automated way and served as the reference values. The study was approved by the Columbia University Irving Medical Center Institutional Review Board and the Nottingham Trent University Ethics Committee.

\subsection{Gradient-Boosted Regression Trees}

Assume $D \coloneqq \{(x_{i},y_{i})\}_{i=1}^{n}$ is the training set where $x_{i}=(x_{i}^{j})_{j=1}^{p} \in X \subseteq \mathbb{R}^{p}$ and $y_{i} \in Y \subset \mathbb{R}$.
Gradient-boosting \cite{Friedman2001Greedy} is a powerful machine learning algorithm that constructs a model $f : X \rightarrow Y$ by relying on stage-wise additive modelling and minimising the expected value of some empirical loss function $L$. The model is obtained through the recursive relationship: $f_0(x) = \gamma$, $\ldots$, $f_t(x) = f_{t-1}(x) + \eta\cdot m_t(x)$. In this formulation, $f_0$ is the base learner, $\gamma$ is an initial approximation, $f_t$ is the model at iteration $t$, $m_t$ represents the weak learner added during iteration $t$ to boost the model, and $\eta$ is the learning rate.

In the case of GBRTs, the most frequently employed $L$ is the mean squared error (MSE), $\gamma$ is chosen as the average outcome of the training instances ($\frac{1}{n}\sum_{i=1}^{n} y_i$), and regression trees represent the weak learners. 
The decision tree at iteration $t$ is chosen to approximate the residual (or else the negative derivative of the loss function with respect to $\hat{y_i} = f_{t-1}(x_i)$), $m_t = \arg\min_m \frac{1}{n}\sum_{i=1}^{n}(-g_{i,t} - m(x_i))^2$, where $g_{i,t} = \frac{\partial L(y_i,\hat{y_i})}{\partial\hat{y_i}}$ is the functional gradient of the $i$-th training instance at iteration $t$.
The decision tree at iteration $t$ recursively creates $M_t$ disjoint regions $\{r_{j}^{t}\}_{j=1}^{M_t}$ through partitioning the instance space. Each of these regions is termed a leaf. The parameter value  $\theta_{t}^{j}$ for leaf $j$ at tree $t$ is commonly determined (given a fixed structure) through a one-step Newton method: $\theta_{t}^{j} = -\frac{\sum_{i\in I_{t}^{j}} g_{t}^{i}}{(\sum_{i\in I_{t}^{j}} h_{t}^{i} + \lambda)}$, where $I_{t}^{j} = \{(x_i, y_i) \: | \: x_i \in r_{j}^{t}\}_{i=1}^{n}$ is the instance set of leaf $j$ for tree $t$, $h_{i}^{t}$ is the second derivative of the $i$-th training instance with  respect to $\hat{y_i}$, and $\lambda$ acts as a regularization parameter. 
Hence, $m_t$ can be denoted as: $m_t(x) = \sum_{j=1}^{M_t} \theta_{t}^{j} \mathds{1}[x \in r_{t}^{j}]$ where $\mathds{1}$ is the indicator function. Lastly, to generate a prediction for a target sample $x_{te}$, the final GBRT model sums up the values of the leaves traversed by $x_{te}$ over all $T$ iterations: $\hat{y}_{te} = \sum_{t=1}^{T} m_t(x_{te})$.

\subsection{Instance-based Uncertainty Quantification}

\begin{figure}
\centering
\includegraphics[width=0.85\textwidth]{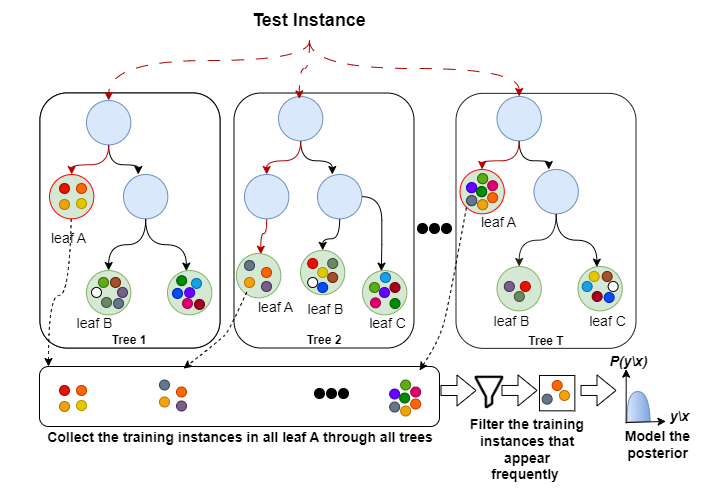}

\caption{IBUG flow chart. For a target instance, IBUG collects the training instances at each leaf it traverses, keeps the $k$ most frequent samples, and then uses those instances to model the output distribution.} \label{fig1}
\end{figure}

The goal is to estimate the conditional probability distribution $P(y|x)$ for some target variable $y$ given some input vector $x$. To allow probabilistic predictions for any GBRT point predictor, we propose to make use of a method that adopts ideas from instance-based learning \cite{Aha1991} and a supervised tree kernel, namely the Instance-Based Uncertainty quantification for GBRTs (IBUG) method \cite{Brophy2022Instance}. To start with, the method capitalises on the fact that GBRTs yield accurate point predictions and uses this scalar output to model the conditional mean in a probabilistic forecast. Next, to further model the conditional output distribution, IBUG uses a supervised tree kernel \cite{davies2014random} to more effectively identify the $k$ training examples with the largest affinity to the target example. In particular, the affinity of the $i$-th training example $x_i$ to a target example $x_{te}$ is given by
\begin{equation} 
A(x_i, x_{te}) = \sum_{i=1}^{T} \mathds{1}[R_t(x_i) = R_t(x_{te})] 
\end{equation}
where $R_{t}(x_{i})$ is the leaf (of the tree $t$) to which $x_{i}$ is assigned. Such a metric uses the structure of the learned trees in the ensemble. Lastly, the method employs the set of those $k$ affinity scores, $A^{(k)}$, to produce a probabilistic prediction. The overall IBUG workflow is illustrated in Fig.~\ref{fig1}. 

Unlike other SOTA methods, IBUG offers numerous choices for modelling the conditional output distribution; the simplest way is, of course, through a normal distribution. In this case, the scalar output of the GBRT model is used to approximate the conditional mean ($\mu_{\hat{y}_{te}} = f (x_{te})$), and then the set $A^{(k)}$ is manipulated to compute the variance $\sigma^{2}_{\hat{y}_{te}}$. To further optimise the calculation of the prediction variance, the following calibration
\begin{equation}
\sigma^2_{\hat{y}_{te}} \leftarrow \gamma \sigma^2_{\hat{y}_{te}} + \delta
\end{equation}
is commonly applied, where $\gamma$ and $\delta$ are tuned on the validation set after the choice of $k$ has been made. The method acts as a wrapper around any GBRT model allowing one to try various GBRT point predictors and then select the model with the best performance. To more flexibly model the output distribution using any parametric or non-parametric distribution, IBUG can use the set $A^{(k)}$ to directly fit (using maximum likelihood estimation) any continuous distribution $D$, including those with high-order moments:
\begin{equation}
\hat{D}_{te} = D\left(A^{(k)} \mid \mu_{\hat{y}_{te}}, \sigma^2_{\hat{y}_{te}}\right).
\end{equation}

Choosing an appropriate value for $k$ is critical for producing accurate probabilistic predictions. In this study, the tuning of $k$ was performed in a held-out validation dataset $D_{\text{val}} \subset D$ using the negative log likelihood (NLL) probabilistic scoring metric. To accelerate the tuning process, through avoiding the repetition of the computationally expensive affinity calculations, the procedure described in Algorithm 3 of \cite{Brophy2022Instance} was adopted, where parameter $\rho$ was used to model instances of abnormally low variance.

\subsection{Implementation} \label{sec:method}

All implementations were in Cython. The experiments were conducted utilising an Intel Core i9 CPU 10900K Comet Lake, 10 Cores, 20 Threads @ 5.3GHz system equipped with 128GB of DDR4 RAM operating @ 2.6GHz. IBUG was applied to XGBoost \cite{XGBoostChenG16}, LightGBM \cite{NIPS2017_LightGBM}, and CatBoost \cite{NEURIPS2018_CatBoost}. We tuned $k$, using values: [3, 5, 7, 9, 11, 15, 31, 61, 91, 121, 151, 201, 301, 401, 501, 601, 701]. The parameters $\gamma$ and $\delta$ were tuned using values ranging from $1 \times 10^{-8}$ to $1 \times 10^{3}$ with additional multipliers $[1.0, 2.5, 5.0]$. The number of trees, $T$, was tuned using values $[10, 25, 50, 100, 250, 500, 1000, 2000]$ (early stopping \cite{pmlr-v119-duan20a} was used for NGBoost). The learning rate was tuned using values $[0.01, 0.1]$. We also optimised: the maximum number of leaves $h$ using values $[15, 31, 61, 91]$, the minimum number of leaves using values $[1, 20]$, and the maximum depth $d$ using values $[2, 3, 5, 7, -1]$ (indicating no limit). The $p$ parameter was adjusted based on the minimum variance obtained from the validation set predictions. We employed 5-fold cross-validation to generate 5 different 80/20 train/test folds. For each fold, the 80\% training set was randomly divided into a 60/20 train/validation set for hyperparameter tuning. Upon tuning the hyperparameters, the model was retrained using the complete 80\% training set. 
To test IBUG's flexibility in posterior modelling, we modelled each probabilistic prediction using the following distributions: normal, skewnormal, lognormal, Laplace, student t, logistic, Gumbel, Weibull, and KDE.

\subsection{Evaluation Metrics}

To evaluate probabilistic performance, the continuous ranking probability score (CRPS), NLL, check score, and interval score were used \cite{Gneiting2007Strictly}. For all metrics, the lower the better. To gauge point performance, the root mean squired error (RMSE), the mean absolute error (MAE), the mean absolute percentage error (MAPE), the R$^{2}$ measure, and correlation were used. IBUG was compared to three recent gradient boosting algorithms that provide probabilistic predictions, namely NGBoost \cite{pmlr-v119-duan20a}, PGBM \cite{Sprangers2021ProbabilisticGB}, and CatBoost with uncertainty (CBU) \cite{malinin2021uncertainty}. 
\subsection{Exemplars}
To facilitate a comprehensive understanding of our results, we provide conditional output distributions and the respective confidence intervals (CIs) for two high and two low accuracy (test set) predictions. 

\subsection{Explainability}

A benefit of using GBRTs is that it is straight-forward to retrieve feature importance scores that indicate how valuable each attribute was in the construction of the model. In this study, feature importance scores were calculated for all models using the 'Gain' metric which quantifies the relative contributions.
\section{Results}
\begin{figure}[H]
    \centering
    
    % First figure on the left side
    \begin{minipage}{0.49\textwidth}
        \centering
        
        % First Row
        \begin{subfigure}[b]{0.75\textwidth}
            \includegraphics[width=\textwidth]{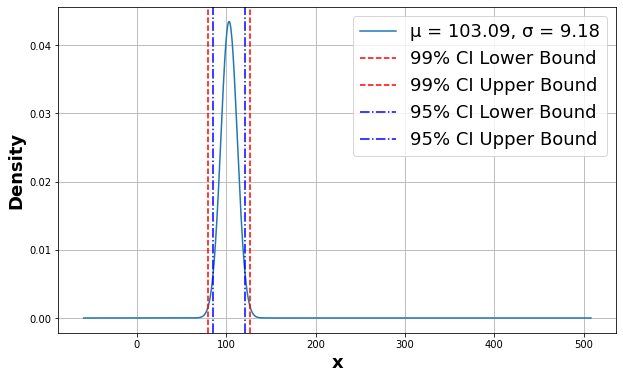}
            \caption{}
            \label{fig:normal2}
        \end{subfigure}
        \begin{subfigure}[b]{0.75\textwidth}
            \includegraphics[width=\textwidth]{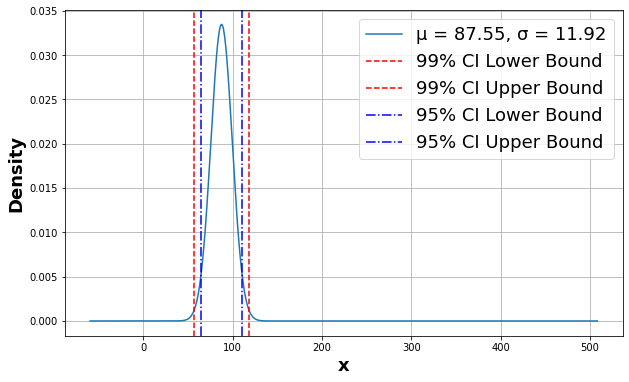}
            \caption{}
            \label{fig:normal9}
        \end{subfigure}
        
        % Second Row
        \begin{subfigure}[b]{0.75\textwidth}
            \includegraphics[width=\textwidth]{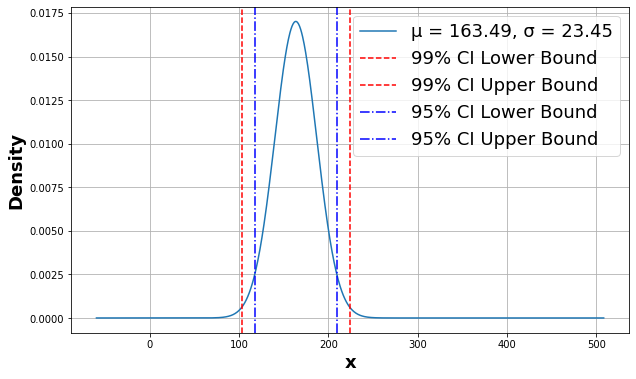}
            \caption{}
            \label{fig:normal8}
        \end{subfigure}
        \begin{subfigure}[b]{0.75\textwidth}
            \includegraphics[width=\textwidth]{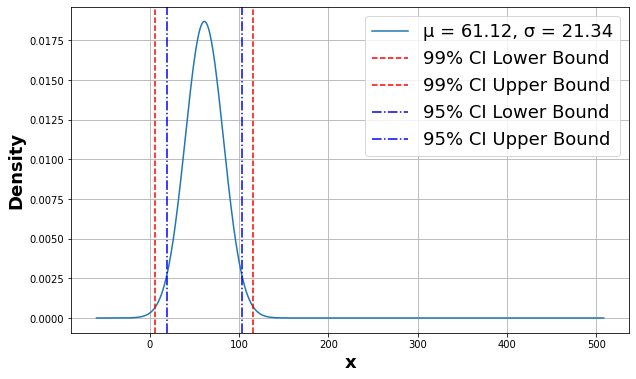}
            \caption{}
            \label{fig:normal14}
        \end{subfigure}
        \caption{The conditional output normal distributions for test instances that were predicted with high [(a) and (b)] and low [(c) and (d)] accuracy.}
        \label{fig:normal_images}
    \end{minipage}%
    % No space to ensure figures are side by side
    \hfill
    \begin{minipage}{0.49\textwidth}
        \centering
        
        % First Row
        \begin{subfigure}[b]{0.75\textwidth}
            \centering
            \includegraphics[width=\textwidth]{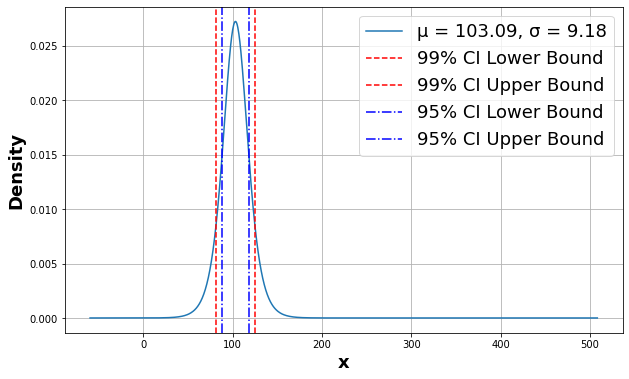}
            \caption{}
            \label{fig:logistic2}
        \end{subfigure}
        \begin{subfigure}[b]{0.75\textwidth}
            \centering
            \includegraphics[width=\textwidth]{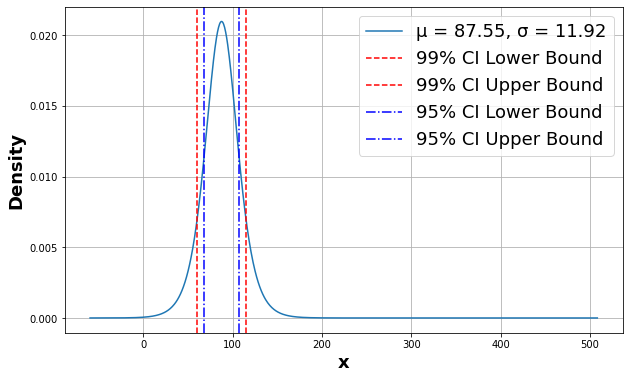}
            \caption{}
            \label{fig:logistic9}
        \end{subfigure}
        
        % Second Row
        \begin{subfigure}[b]{0.75\textwidth}
            \centering
            \includegraphics[width=\textwidth]{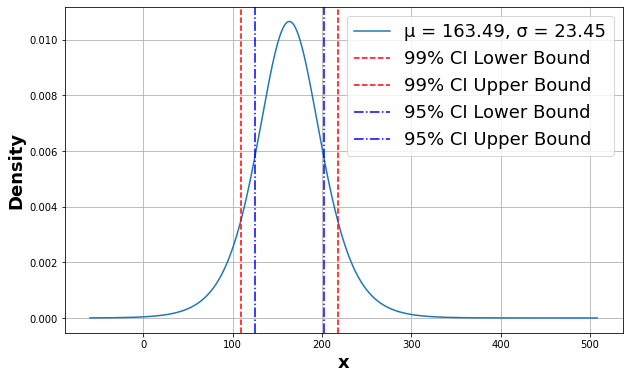}
            \caption{}
            \label{fig:logistic8}
        \end{subfigure}
        \begin{subfigure}[b]{0.75\textwidth}
            \centering
            \includegraphics[width=\textwidth]{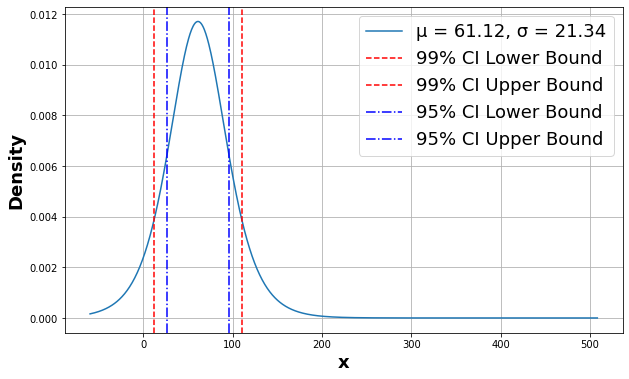}
            \caption{}
            \label{fig:logistic14}
        \end{subfigure}
        
        \caption{The conditional output logistic distributions for test instances that were predicted with high [(e) and (f)] and low [(g) and (h)] accuracy.}
        \label{fig:logistic_images}
    \end{minipage}
\end{figure}

\begin{figure}[H]
\centering
\includegraphics[width=0.7\textwidth]{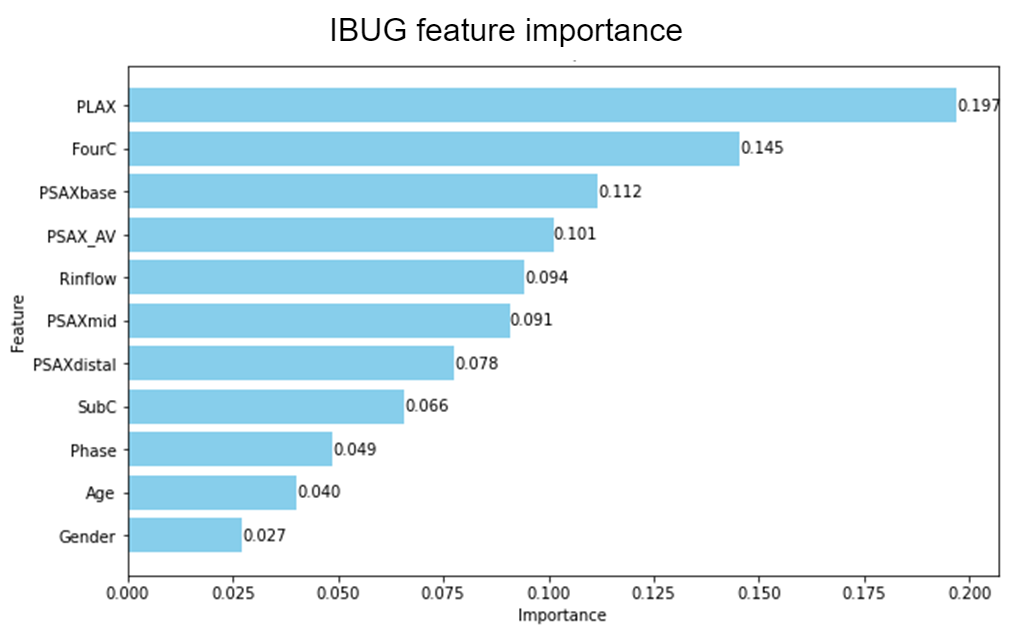}

\caption{Feature importance plot for the IBUG model with CatBoost as the base learner.} \label{fig3}
\end{figure}

The final set of hyperparameters for each method and the corresponding tuning and training times are listed in Table \ref{tab1}. Table \ref{tab2} compares the probabilistic performance of the IBUG model against the three SOTA probabilistic prediction methods. IBUG model with CatBoost as the base learner provided the lowest average scores in all CRPS, NLL, Check Score and Interval Score indices. In Table \ref{tab3}, the point performance of all methods is provided. In overall, IBUG method with CatBoost base learner displayed the best performance once again. Table \ref{tab4} shows the importance of variance calibration in the probabilistic performance of IBUG. Table \ref{table5} demonstrates that the logistic (parametric) distribution better fits the underlying data than assuming normality. 
In Fig. \ref{fig:normal_images} and Fig. \ref{fig:logistic_images}, we illustrate the conditional output distributions for four representative test cases (two that were predicted with high accuracy and two that were predicted with low accuracy), when normal and logistic probabilistic density functions were used for modelling, respectively.
Table \ref{tab:Normal_Logistic_pdf} lists the 95\% and 99\% confidence intervals for the above cases.
These results showcase the appropriateness of the proposed framework for providing uncertainty scores for RV volume predictions.
Lastly, in Fig. \ref{fig3}, we illustrate the "Gain" feature importance score for all eleven features in the best IBUG model. The parasternal long axis (PLAX), four chambers (Four C) and parasternal short axis at base level (PSAX Base) standard views were the top three contributors to the model predictions.

\begin{table}[H]
\centering
\caption{The final set of hyperparameters used for each method. Also shown are the tuning and training times. IBUG was applied to CatBoost, XGBoost, and LightGBM.}
\label{tab1}
\begin{tabular}{|c|c|c|c|c|c|c|}
\hline
Parameter & CatBoost & XGBoost & LightGBM & NGBoost & PGBM & CBU \\
\hline
K & 5 & 15 & 3 & - & - & - \\
$\delta$ & 1 & 0.5 & 0.5 & 5 & 10 & 1 \\
Operation & add & mult & mult & add & add & add \\
min scale & 6.164 & 13.826 & 2.055 & - & - & - \\
n estimators (trees) & 100 & 25 & 25 & 244 & 250 & 250 \\
maximum depth & 5 & 2 & -1 & - & - & - \\
learning rate & 0.1 & 0.1 & 0.1 & 0.01 & 0.01 & 0.1 \\
minimum data in\_leafv & 1 & - & - & - & 20 & 1 \\
minimum child weight & - & 20 & 20 & - & - & - \\
number of leaves & - & - & 15 & - & 15 & 15 \\
max bin & 255 & 255 & 255 & 255 & 255 & 255 \\
tune+train time (s) & 67.369 & 19.932 & 14.260 & 5.370 & 872.663 & 81.929 \\
\hline
\end{tabular}
\end{table}

\begin{table}[H]
\centering

\caption{Probabilistic performance comparison on the test set (five folds). IBUG results are for the case when CatBoost was the base learner. IBUG results have been averaged over all nine posterior output distributions.}
\label{tab2}
\begin{tabular}{|c|c|c|c|c|}
\hline
Method & NLL & CRPS & Check Score & Interval Score \\
\hline
\textbf{IBUG} & \textbf{4.747} & \textbf{15.398} & \textbf{7.775} & \textbf{73.380} \\
NGBoost & 7.571 & 22.174 & 11.177 & 141.618 \\
PGBM & 6.136 & 20.796 & 10.492 & 122.401 \\
CBU & 5.780 & 19.524 & 9.853 & 110.140 \\
\hline
\end{tabular}
\end{table}

\begin{table}[H]
\centering
\caption{Point performance comparison on the test set (five folds). IBUG results are for the case when CatBoost was the base learner.}
\label{tab3}
\begin{tabular}{|c|c|c|c|c|c|}
\hline
Method & MAE & RMSE & MAPE  & R$^{2}$ & Correlation \\
\hline
\textbf{IBUG} & \textbf{22.75} & \textbf{26.292} & \textbf{20.22} & \textbf{0.666} & \textbf{0.824} \\
%XGBoost & 28.205 (7.674) & 34.322 (5.066) & 23.869 (6.706) & 0.431 & 0.175 \\
%LightGBM & 28.275 (7.294) & 32.618 (4.739) & 25.891 (6.890) & 0.486 & 0.742 \\
NGBoost & 28.114 & 32.269 & 24.406 & 0.496 & 0.736 \\
PGBM & 27.479  & 31.27 & 25.147 & 0.527 & 0.768 \\
CBU & 26.378  & 30.127  & 22.974  & 0.561 & 0.772 \\
\hline
\end{tabular}
\end{table}

\begin{table}[H]
\centering
\caption{Probabilistic performance comparison of IBUG method with and without variance calibration. IBUG results are for the case when CatBoost was the base learner.}
\begin{tabular}{|c|c|c|c|c|}
\hline
Operation & NLL & CRPS & Check Score & Interval Score  \\
\hline
\textbf{With Calibration} & \textbf{4.747} & \textbf{15.398} & \textbf{7.775} & \textbf{73.38}  \\
Without Calibration & 4.781 & 15.457 & 7.805 & 74.044  \\
\hline
\end{tabular}
\label{tab4}
\end{table}

\begin{table}[H]
\centering
\caption{Probabilistic performance comparison when assuming normal and logistic distributions for modelling the underlying data.}
\begin{tabular}{|c|c|}
\hline
  Distribution & NLL \\
\hline
Normal & 5.10466  \\
\textbf{Logistic} & \textbf{5.00837} \\
\hline
\end{tabular}
\label{table5}
\end{table}

\begin{table}[H]
    \centering
    \caption{The 95\% and 99\% Confidence Intervals for both Normal and Logistic distributions for four representative test set cases, two that were predicted with high accuracy (low APE) and two that were predicted with low accuracy (high APE).}
    \begin{tabularx}{\linewidth}{|C|C|C|C|C|C|C|}
        \hline
        Prediction & \centering APE (\%) & Point & \multicolumn{2}{|c|}{\centering Normal Distribution} & \multicolumn{2}{c|}{\centering Logistic Distribution} \\
        \cline{4-7}
        \centering  Accuracy &  & {\centering  Prediction} & {\centering 95\% CI} & {\centering 99\% CI} & {\centering 95\% CI} & {\centering 99\% CI} \\
        \hline
        High & 3.090 & 103.090 & 35.979 & 47.286 & 30.196 & 42.697 \\
              & 0.508 & 87.553  & 46.739 & 61.428 & 39.227 & 55.466 \\
        \hline
        Low & 10.169 & 163.492 & 91.932 & 120.825 & 77.157 & 109.099 \\
            & 10.119 & 61.119  & 83.650 & 109.939 & 70.206 & 99.270 \\
        \hline
    \end{tabularx}
    \label{tab:Normal_Logistic_pdf}
\end{table}

\section{Discussion and Conclusions}

In this paper, we presented an uncertainty quantification framework when predicting RV volumes through the use of GBRTs on 2DE tabular data (such as area measurements, age, gender and cardiac phase information). To get an estimate of the prediction uncertainty, we employed the IBUG method which uses the learned tree structure to identify the $k$ nearest training samples to a target instance. The results on a small-scale dataset demonstrate that this simple wrapper yields improved probabilistic and point performances over other SOTA methods. 
The appropriateness of the proposed framework for providing uncertainty scores for RV volume predictions was showcased by providing conditional output distributions and confidence intervals for four exemplar cases. Additional research is required, involving a larger sample size of patients and encompassing a broader range of RV volumes, to substantiate these findings.
The estimated uncertainty embodies both aleatoric and epistemic types of uncertainty since IBUG is an instance-based approach and also predictions on the training set were used to tune $k$, $\gamma$, and $\delta$. Overfitting was observed in IBUG's point performance which is a typical finding when the size of the dataset is small.
This work aligns with trustworthy artificial intelligence \cite{Li2023Trustworthy} since it can be used to enhance the decision-making process and reduce risks. It could help overcome mistrust which is a major barrier to the deployment of machine learning systems in the clinical setting. The calculated feature importance scores can be used for reducing the number of required 2DE views, which in turn could enhance the proposed pipeline's clinical application.
\clearpage

\acknowledgements

\section*{Funding}

Tuan Aqeel Bohoran is funded by the European Union's Horizon 2020 research and innovation programme under the Marie Sklodowska-Curie grant agreement No 801604. 

\bibliographystyle{spiebib}
\bibliography{bibliography}

% This table is required by ICMV for review and will not be published

\clearpage

\section*{AUTHORS' BACKGROUND}

\begin{table}[h]
    \centering
    \begin{tabular}{|c|p{22mm}|p{45mm}|p{51mm}|}
        \hline
        Name & Title & \multicolumn{1}{c|}{Research Field} & \multicolumn{1}{c|}{Personal website} \\
        \hline
        Tuan Aqeel Bohoran & Researcher & Data Science, Computer Science, Machine Learning, Deep Learning, Medical Imaging, Mathematics, Computer Vision, Cardiovascular MRI & \url{https://github.com/tuanaqeelbohoran} \\
        \hline
        Polydoros N. Kampaktsis & Assistant Professor & Valvular Disease, Coronary Artery Disease, transcatheter therapies, artificial intelligence, cardiovascular imaging & \url{https://www.columbiacardiology.org/profile/polydoros-kampaktsis-md} \\
        \hline
        Laura McLaughlin & Medical Student &  & \url{https://www.linkedin.com/in/laura-mclaughlin-4112ab82}\\
        \hline
        Jay Leb & Assistant Professor & Cardiovascular Disease, CT, MRI & \url{https://www.columbiaradiology.org/profile/jay-s-leb-md}\\
        \hline
        Serafeim Moustakidis & Co-founder/partner & Healthcare, Energy, Agriculture, Environment, Manufacturing, Security and Safety, Construction and building technology, Infrastructure and Transportation & \url{https://www.aideas.eu/serafeimmoustakidis}\\
        \hline
        Gerry P. McCann & Full Professor & Cardiac Imaging, Aortic Stenosis, Diabetes and Heart Failure, CAD, Clinical Imaging & \url{https://le.ac.uk/people/gerry-mccann}\\
        \hline
        Archontis Giannakidis & Senior Lecturer & Data Science, Computer Science, Machine Learning, Deep Learning, Medical Imaging, Mathematics, Computer Vision, Cardiovascular MRI & \href{https://www.ntu.ac.uk/staff-profiles/science-technology/archontis-giannakidis}{\url{https://www.ntu.ac.uk/staff-profiles/science-technology/archontis-giannakidis}} \\
        \hline
    \end{tabular}
\end{table}

\end{document}